# The role of oxygen vacancies in SrTiO$_3$ at the LaAlO$_3$/SrTiO$_3$ interface

Alexey S. Kalabukhov[1*], Robert Gunnarsson[1], Johan Börjesson[2], Eva Olsson[2], Tord Claeson[1] & Dag Winkler[1]

[1] *Quantum Device Physics Laboratory, Department of Microtechnology and Nanoscience (MC2), Chalmers University of Technology, Göteborg, Sweden*

[2] *Microscopy and Microanalysis, Department of Applied Physics, Chalmers University of Technology, Göteborg, Sweden*

* Electronic address: ascry@chalmers.se

**Strontium titanate, SrTiO$_3$, a widely used substrate material for electronic oxide thin film devices, has provided many interesting features. In a combination with a similar oxide material, LaAlO$_3$, it has recently received great interest. It was suggested that two-dimensional electron gas is formed at the interface between SrTiO$_3$ and LaAlO$_3$, resulting in high electrical conductivity and mobility. In this report we demonstrate that the transport properties in those heterostructures are very sensitive to the deposition parameters during thin film growth. Using cathode- and photoluminescence studies in conjunction with measurements of electrical transport properties and microstructure we show that the electronic properties observed at a LaAlO$_3$/SrTiO$_3$ interface can be explained by oxygen reduced SrTiO$_3$. In addition, we demonstrate that oxygen can be pushed in and out of the sample, but that re-oxygenation of an initially oxygen depleted LaAlO$_3$/SrTiO$_3$ heterostructure is partly prevented by the presence of the film.**

Strontium titanate, $SrTiO_3$, is a very versatile material. It is a cubic perovskite-structured dielectric material with a wide band-gap of about 3.2 eV at 300 K in its stoichiometric composition [1,3]. Its transparency and dielectric properties have made $SrTiO_3$ one of the most frequently used single crystal substrates for electronic oxide thin film devices. A prominent feature of $SrTiO_3$ is the possibility to control its surface composition at the atomic level. Singly terminated $SrTiO_3$ is an excellent choice as a starting block for growth of atomically sharp oxide heterointerfaces [4,5]. The possibility of charge discontinuity between $SrTiO_3$ and another similar perovskite, $LaAlO_3$, has recently been proposed [6]. Uncompensated charge may appear at the interface because of different formal valences in the ionic limit of Ti (4+) and La (3+) in these materials. As a result, highly interesting electrical properties like the formation of a two-dimensional electron gas was suggested. However, the properties of $SrTiO_3$ itself can easily be modified with a small compositional change. By replacing only a small fraction of Sr with an alkaline earth metal (like, e.g., Ba or Ca) the material can become ferroelectric with $T_C$ up to 136 K [7]. A small addition of Nb, La or Ta makes the material highly (n-type) conducting with a charge carrier concentration of $10^{19}$ $cm^{-3}$ [8,9] and superconducting below 0.4 K [10].

Another method to modify the material properties is to reduce the stoichiometric $SrTiO_3$ and introduce oxygen deficiencies ($\delta$ in $SrTiO_{3-\delta}$). There are mainly three different ways to create oxygen deficient $SrTiO_3$. One way is to anneal stoichiometric crystals at high temperature (800ºC – 1200ºC) in vacuum[9] or in the presence of titanium or hydrogen[11]. Alternatively, reduced $STO_{3-\delta}$ films can be deposited at low oxygen pressure, hence creating inherently oxygen depleted $SrTiO_{3-\delta}$ [12]. It has also been shown that it is possible to make oxygen deficient layers of $SrTiO_{3-\delta}$ by Ar-ion bombardment[13,15].

At a glance, the transport measurement data from $LaAlO_3/SrTiO_3$ interfaces show striking similarities with corresponding data from oxygen depleted $SrTiO_{3-\delta}$. The temperature dependence of resistivity and charge carrier density is similar, and, as we will demonstrate in this paper, $LaAlO_3/SrTiO_3$ show cathode luminescence, similar to what has been reported in $SrTiO_{3-\delta}$ [15]. Furthermore, it has been reported that the high charge carrier density ($n_s \sim 10^{17}$ cm$^{-2}$) found by Ohtomo *et al*[6] only can be obtained under specific low-pressure (about $10^{-6}$ mbar) deposition conditions[6,16,17]. In this paper we provide experimental data indicating that the conductivity in $LaAlO_3/SrTiO_3$ heterostructures is due to oxygen reduced $SrTiO_{3-\delta}$ substrate, and that the oxygen vacancies are introduced during the deposition process. We also show that the environmental conditions *alone* are sufficient to reduce the substrate and obtain the highly conducting layer.

The experimental details are described below. Ultra-thin $LaAlO_3$ films were grown on (100) $SrTiO_3$ substrates by pulsed laser deposition (PLD). Prior to deposition the $SrTiO_3$ substrates were chemically and thermally treated in order to obtain smooth single terminated ($TiO_2$) surfaces[4,5]. The atomically flat character of the substrate surface was confirmed by atomic force microscopy, where we could observe atomic steps with a height of about 4 Å corresponding to one unit cell of $SrTiO_3$. A time chart of the deposition parameters is shown in Fig. 1a. The substrates were attached to a heater block using silver glue and heated to 70ºC at ambient pressure for 15 min, then introduced into vacuum ($10^{-7}$ mbar) and heated further at a rate of 10ºC/min. At 250ºC, oxygen was let into the chamber to a pressure of $10^{-4}$ mbar and the temperature was thereafter increased by 40ºC/min up to 800ºC. Two scenarios were then used. (*i*) At 800 ºC ultra-thin film of $LaAlO_3$ was deposited, or (*ii*) the substrate was kept at 800ºC for a time corresponding to the film deposition. Two different oxygen pressure regimes were used, "low pressure" ($10^{-6}$ mbar) and "high pressure" ($10^{-4}$ mbar), as shown in Fig. 1a. The stability at low pressure was limited by the precision of the mass-flow controller, but was at all times better than $\pm 2.5 \times 10^{-7}$ mbar. A KrF excimer laser (pulse width ~20 ns, repetition rate 1 Hz, energy density 2 J/cm$^2$) was focused onto a LaAlO$_3$ single crystal, which was used as a target for the ablation. *In-situ* reflection high-energy electron diffraction (RHEED) was used to monitor the film growth and the surface morphology during the deposition process. RHEED intensity and pattern observed in the growth process are shown in Fig. 1b and 1c, respectively. The well defined and slightly damped RHEED oscillations indicate an atomic layer-by-layer growth[5]. The growth rate was monitored by the number of oscillations – about 15 pulses were required for the completion of one unit cell (uc) of LaAlO$_3$. The thickness of the LaAlO$_3$ films was varied in the range of 7 to 35 unit cells. Bright and spotty RHEED pattern (Fig. 1c) were observed even after deposition of 35 uc layers LaAlO$_3$ film which indicates a very smooth film surface. This was also confirmed by atomic force microscopy. After deposition, samples were cooled down at the rate of 10 ºC/min keeping the same oxygen pressure as during deposition. For some process runs, the films were oxidized during cooling in 500 mbar oxygen pressure.

Two additional methods were used to reduce the oxygen content in SrTiO$_3$ substrates: *(i)* Ar$^+$-ion irradiation with an acceleration voltage of 300 eV, and beam current of 0.2A/cm$^2$, for 10 min and *(ii)* vacuum annealing at a pressure of 10$^{-7}$ mbar at 800ºC for 30 min.

After the deposition at low oxygen pressure (10$^{-6}$ mbar), the samples changed color, from transparent (white) to grey hue, which is characteristic for oxygen reduced SrTiO$_3$. Since oxygen deficient SrTiO$_3$ is also known to show cathode luminescence, we irradiated as-deposited heterostructures with an electron beam and observed strong blue light emission, see Fig. 2. For cathode luminescence we used the *in-situ* RHEED e-gun with an acceleration voltage of 35 kV and a beam current of 50 µA. The samples were kept at room temperature. Photos of the emitted light (Fig. 2 and Fig. 5) were captured by professional digital camera. The luminescence was usually bright enough to be observed by the naked eye, but in some cases the intensity was quite weak. Therefore photos were taken with exposure times varying between 2 s and 30 s. To compensate for the different exposure times and keep the information in the images correct the pictures shown here were light enhanced in such way that the intensities are comparable.

For the low oxygen pressure samples, the intensity of the light was comparable independent of whether there was a film on top of the substrate or not (Fig 2 a,b). Cathode luminescence, though with weaker intensity, was also observed from films produced at higher oxygen pressure ($10^{-4}$ mbar), see Fig 2 c,d. The light has the same color from both heterostructures, the reduced $SrTiO_{3-\delta}$ substrates, and the $Ar^+$-ion etched $SrTiO_3$ (Fig. 2e). Cathode luminescence emitted from a $LaAlO_3$ substrate (Fig. 2f) is more whitish in color than from $SrTiO_3$, and the luminescence became even more white after $LaAlO_3$ had been oxygen reduced. We also note that after $SrTiO_3$ samples were annealed in 500 mbar oxygen at 600 ºC the cathode luminescence was too weak to be observed by the naked eye, this is further discussed below.

Hence, blue-light cathode luminescence was observed in both $LaAlO_3/SrTiO_3$ heterostructures and intentionally oxygen depleted $SrTiO_3$ (i.e. Ar-ion etched), as well as in $SrTiO_3$ substrates treated in the same way and using exactly the same conditions as during film deposition (except for the actual deposition itself). As $LaAlO_3$ shines with a different color we may conclude that the luminescence is not produced in the film itself. In addition, we found that both the high and the low pressure conditions are sufficient for the substrate to become cathode luminescent.

To quantify the wavelength of the emitted light we conducted room-temperature ultraviolet photoluminescence experiments using a 350 nm argon laser. Two samples with different $LaAlO_3$ film thicknesses (7 uc and 15 uc) and an $Ar^+$ etched $SrTiO_3$ were studied. The results of the photoluminescence measurements are presented in Fig. 2g.

Within the precision of the experiment the emitted light from both LaAlO$_3$/SrTiO$_3$ heterostructures and doped SrTiO$_3$ substrate has the same peak in wavelength of 460 nm. The position of the photoluminescence peak agrees well with the wavelength of light emitted from Ar$^+$ bombarded SrTiO$_3$ [15], where it is ascribed to oxygen vacancies creating states within the bandgap. The light emitted from samples processed at high-pressure conditions was too weak to be detected using our photoluminescence equipment.

However, we would like to emphasize that the cathode luminescence was observed in LaAlO$_3$/SrTiO$_3$ heterostructures, as well as in reduced SrTiO$_3$ substrates, prepared in *both* high and low oxygen pressure. We therefore conclude that oxygen vacancies can be formed in the SrTiO$_3$ substrate without the deposition of LaAlO$_3$ films, even at high oxygen pressure. Our cathode- and photo-luminescence results are in good agreement with previous reports on luminescence in oxygen depleted SrTiO$_3$ [18], where the cathode luminescence was observed only in oxygen reduced strontium titanate. Also, the light intensity depends strongly on the electrical conductivities of the reduced SrTiO$_3$, but the shape of the emission band does not depend on the particular doping level. This indicates that the light originates from oxygen reduced regions in the substrate.

Electrical transport measurements of LaAlO$_3$/SrTiO$_3$ heterostructures and reduced SrTiO$_3$ substrates were made in a four point van der Pauw configuration[19] in the temperature range 2 K – 300 K and in magnetic field up to 5 T. Gold pads were defined by sputtering using a Ti adhesion layer to make contacts to the samples. Several reports have already been presented in which the electrical transport properties of LaAlO$_3$/SrTiO$_3$ thin films and oxygen depleted SrTiO$_3$ have been carefully described[6,9,17]. The temperature dependences of the sheet resistance $R_{xx}$, the Hall mobility $\mu_H$ and the charge carrier density $n_s$ are presented in Fig. 3. The behavior of our samples agree very well with the general temperature dependence presented by, e.g.,

Ohtomo *et al.* [6] on LaAlO$_3$/SrTiO$_3$ films and Kan *et al.* [15] and Reagor *et al.* [14] on Ar-ion irradiated SrTiO$_3$.

The general behavior is described next. The sheet resistance drops with temperature and saturates at around 10 K. The low temperature resistance value is about $10^{-2}$ Ω/□ for the LaAlO$_3$/SrTiO$_3$ films made at low oxygen pressure and oxygen reduced SrTiO$_3$. Oxidized films showed much higher resistivity, around 1 k Ω /□ at low temperature. The same values of resistance were found for LaAlO$_3$/SrTiO$_3$ films grown at high oxygen pressure. Additionally, a SrTiO$_3$ substrate treated in deposition conditions at $10^{-4}$ mbar oxygen pressure and 800°C did not show any conductivity. However, we did observe cathode luminescence from the SrTiO$_3$ substrate treated at high pressure ($10^{-4}$ mbar) conditions. We suggest two possible reasons for this apparent contradiction. Either it could be due to clustering of oxygen vacancies, which has previously been observed [20] in SrTiO$_{3-\delta}$. An alternative explanation could be that the cathode luminescence has a different origin in this sample than in the low-pressure prepared samples. This cannot be ruled out without information from spectroscopy.

The Hall mobilities deduced from the resistivity and Hall measurements are the same for the LaAlO$_3$/SrTiO$_3$ films and the oxygen depleted SrTiO$_3$ substrate ($10^4$ cm$^2$V$^{-1}$s$^{-1}$). However, when the sample is annealed in oxygen the Hall mobility is suppressed by one order of magnitude and the resistance increases by five orders of magnitude. Similar behavior is seen for LaAlO$_3$/SrTiO$_3$ deposited at high oxygen pressure. Further oxidation of the films at 600 °C in 500 mbar oxygen had only a minor influence on the sheet resistivity and the Hall mobility.

From the transport measurements we conclude that independently on whether the oxygen depletion was produced by Ar-ion milling, vacuum annealing, or during LaAlO$_3$ deposition at low oxygen pressure, the sheet resistivity, charge carrier concentration and Hall mobility have the same temperature dependences. This behaviour is strikingly similar to that reported previously for reduced $SrTiO_{3-\delta}$: annealed [8,9], doped by Nb or La [8] or oxygen depleted by $Ar^+$ bombardment[13,14]. The congruence of electrical properties suggests that high electrical conductivity and mobility observed in $LaAlO_3/SrTiO_3$ heterostructures are due to oxygen vacancies in $SrTiO_3$. Although the temperature dependence of the electrical transport properties is similar for $LaAlO_3/SrTiO_3$ heterostructures produced at high and low oxygen pressure, the resistivity is different by almost five orders of magnitude and the mobility is different by one order of magnitude. Hence, re-oxygenation of the $LaAlO_3/SrTiO_3$ heterostrucure fabricated at low oxygen pressure has the same effect on the transport properties as deposition at high pressure.

We observed conductivity in $LaAlO_3/SrTiO_3$ heterostructures deposited at high oxygen pressure ($10^{-4}$ mbar). However, conductivity was *not* observed in a $SrTiO_3$ substrate treated at the same high-pressure condition. This suggests that the conductivity is not due to the substrate preparation alone, but also may be due to the effect of the film deposition or the interface properties. This issue will be further addressed below.

Before this, we should also discuss our observation of conductivity in $SrTiO_3$ substrates annealed in low-pressure ($10^{-6}$ mbar) deposition conditions. This is in contrast to other reports[6,17], where substrates treated at deposition conditions did not show any conductivity. However, similar low sheet resistance ($5 \times 10^{-3} \Omega/\square$) was recently reported in $SrTiO_3$ annealed at a low oxygen pressure of $5 \times 10^{-5}$ torr at 1000°C [21], which agrees with our data. They suggest that if the pressure is low enough, like $10^{-6}$ mbar as for typical high-mobility $LaAlO_3/SrTiO_3$ heterostructure depositions, even 800°C is sufficient to reduce the $SrTiO_3$. The properties of the $LaAlO_3/SrTiO_3$ interface are very sensitive to the actual environmental conditions regarding temperature and pressure. Thus minor differences in the actual procedures possibly can explain the absence of conductivity in Refs. 6 and 17.

A critical question has been raised regarding the exceptionally high charge carrier density observed in the LaAlO$_3$/SrTiO$_3$ heterointerface [6,17]. We observe n$_s$ ~ 10$^{16}$ – 10$^{17}$ cm$^{-2}$. However, we stress that our data suggest that the charge carriers are not localized to the interface alone but is more a bulk effect, as was suggested also by Siemons *et al.*[17].

We also measured the sheet resistance as function of magnetic field. However, the magnetic field we applied to the sample (up to 5 T, perpendicular to the sample) was not sufficient to observe Shubnikov – de Haas oscillations. This type of oscillation is well known to occur in conducting SrTiO$_3$ [22], and has been reported also in structures of oxide thin films grown on SrTiO$_3$ substrates [6,16].

In order to understand the microscopic origin of these effects in the LaAlO$_3$/SrTiO$_3$ systems we performed high-resolution transmission electron microscopy (TEM) using a Philips CM 200 field emission gun TEM operating at 200kV. The TEM samples were prepared using standard techniques including tripod polishing and ion milling.

Fig. 4 shows high-resolution TEM cross-section micrographs of the 15 uc thick LaAlO$_3$ film deposited at an oxygen pressure of 10$^{-6}$ mbar. The incident electron beam was parallel to the [100] axis of the SrTiO$_3$ substrate. The orientation relationship between LaAlO$_3$ and SrTiO$_3$ is [100]$_{LAO}$//[100]$_{STO}$ and [010]$_{LAO}$//[010]$_{STO}$. In Fig. 4a, a U-shaped dark contrast can be seen in the SrTiO$_3$ substrate near the film/substrate interface reaching up into the film at several positions and coinciding with misfit dislocations at the interface. The average distance between these dislocations is approximately 15 nm. Fig 4b shows the SrTiO$_3$/LaAlO$_3$ interface between the dislocations with higher resolution, and Fig 4c shows a fast Fourier transformed filtered image of Fig. 4b. A clear coherence of the film/substrate interface can be seen in Fig. 4b and Fig. 4c. Hence, the TEM investigation of the interface shows the presence of misfit dislocations. Considering the orientation relationship between the LaAlO$_3$ and the SrTiO$_3$ ([100]$_{LAO}$//[100]$_{STO}$ and [010]$_{LAO}$//[010]$_{STO}$) and the corresponding lattice mismatch of 3 %, the expected distance between dislocations for a fully relaxed film is 11 nm. The observed distance is 15 nm and the overall film is thus not completely relaxed. The LaAlO$_3$ unit cell is smaller than the SrTiO$_3$ and the residual strain in the SrTiO$_3$ at the interface will therefore be compressive along the [100] and [010] directions. In addition, the contrast observed in the TEM images in the vicinity of the dislocations shows that they give rise to local strain fields in the SrTiO$_3$ reaching about 10 nm into the SrTiO$_3$ substrate. It is known that crystalline defects in SrTiO$_3$ can enhance the diffusion of oxygen [23]. The presence of the residual compressive strain in SrTiO$_3$ can also enhance oxygen diffusion during growth of the LaAlO$_3$ films. This can explain that SrTiO$_3$ substrates annealed at high oxygen pressure are not conducting, in contrast to LaAlO$_3$/SrTiO$_3$ heterostructures. Based on the perovskite tolerance factor, *t*, we can estimate the activation energy for oxygen vacancy diffusion. For SrTiO$_3$ (*t*=0.8072) it is ~ 0.75 eV, and for LaAlO$_3$ (*t*=0.8857) we find ~ 2.2 eV [24]. Hence, it is reasonable to believe that during the initial growth of LaAlO$_3$ on SrTiO$_3$ in a low pressure environment the oxygen is more easily removed from SrTiO$_3$ than transported through LaAlO$_3$.

The SrTiO$_3$ is also known to be piezoresistive. The resistivity decreases for a compressive stress along the <100> axes at temperatures below 110 K [8]. This could explain the small difference in the resistivity values between low-pressure annealed SrTiO$_3$ and LaAlO$_3$/SrTiO$_3$ heterostructures. These arguments suggest that oxygen vacancies in SrTiO$_3$ are responsible for the conductivity in LaAlO$_3$/SrTiO$_3$ heterostructures prepared even at high oxygen pressures. However, at this stage we can not exclude the possibility that polar discontinuity between LaAlO$_3$ and SrTiO$_3$ could have a minor effect on the electrical properties.

We have also monitored the cathode luminescence in a LaAlO$_3$/SrTiO$_3$ sample when subject to different annealing processes, see Fig. 5. The luminescence from the as deposited sample is intense. However, after exposure to 0.5 bar oxygen at 600ºC for 2 hrs the intensity is considerably decreased. In order to reduce the sample we then annealed it *in-situ* for 2 hrs in vacuum (<10$^{-6}$mbar) at 600 ºC for 2hrs. The intensity of the cathode luminescence was then recovered, at least partially, see Fig 5c. The sample was then brought ex-situ to a furnace and kept in a strongly reducing H$_2$/Ar environment at 700ºC for 17 hrs. However, after this procedure the intensity of the luminescent light (Fig. 5d) was about the same as in Fig 5c. The presence of a remanent cathode luminescence can be explained by strain. Strain may prevent re-entrance of oxygen during oxidation of the LaAlO$_3$/SrTiO$_3$ heterostructure, and since the strain is more pronounced in the vicinity of the interface this may explain why LaAlO$_3$/SrTiO$_3$ cannot be made fully insulating by the oxidation. Possibly it could also explain why the reduction/annealing of the sample first in vacuum and then in H$_2$/Ar did not fully remove the oxygen and recover the cathode luminescence intensity.

In summary, we have shown that high electrical conductivity and mobility values of LaAlO$_3$/SrTiO$_3$ heterointerfaces produced at low oxygen pressure are due to oxygen vacancies in a reduced SrTiO$_3$ substrate. The question whether the conductivity of the LaAlO$_3$/SrTiO$_3$ interface deposited at higher oxygen pressure is also due to oxygen vacancies in SrTiO$_3$ or not still remains. It could be a result of stress induced in SrTiO$_3$ or an effect of polar discontinuity. Further investigation is necessary to answer this question.

**Acknowledgements**

We thank J.-O. Yxell for help with the photos of the cathode-luminescence, and Q. X. Zhao for assistance with the photo-luminescence experiments. We also acknowledge communications with the Stanford group and H. Y. Hwang. The work is supported by the Swedish Research Council, Foundation of Strategic Research, and Royal Academy of Sciences, the K.A. Wallenberg foundation and the ESF THIOX.

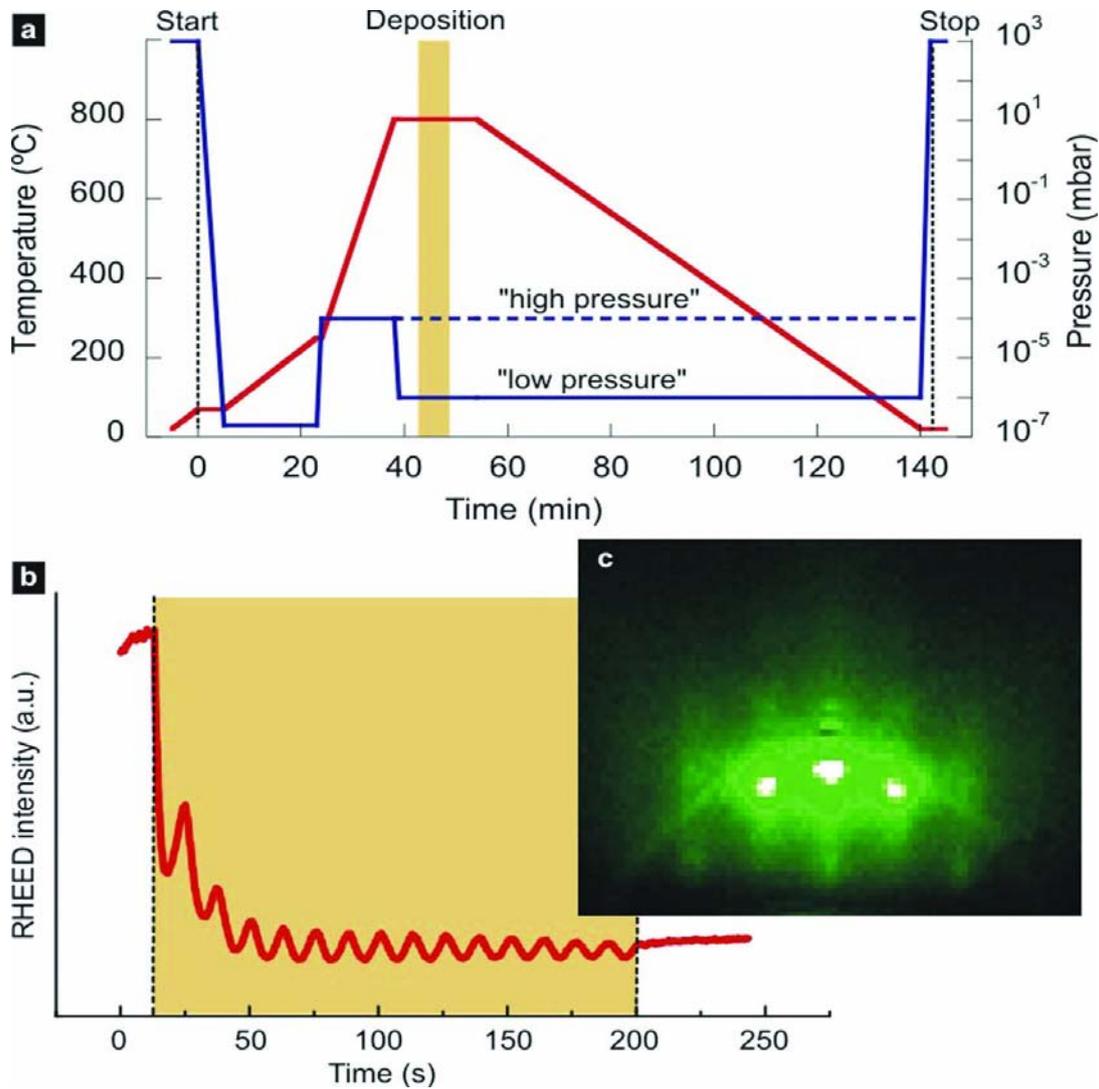

Figure 1. (a) Variation of substrate temperature (red) and chamber pressure (blue) during a deposition process. The epitaxial thin film is formed (at time marked yellow) at high temperature (800°C) and in low-pressure ($10^{-6}$ mbar) flow of oxygen. The deposition chamber is ventilated with argon when the sample has been cooled down to room temperature. (b) RHEED intensity oscillations during growth of 15 unit cells of $LaAlO_3$ film on $SrTiO_3$. (c) RHEED pattern of the film after deposition.

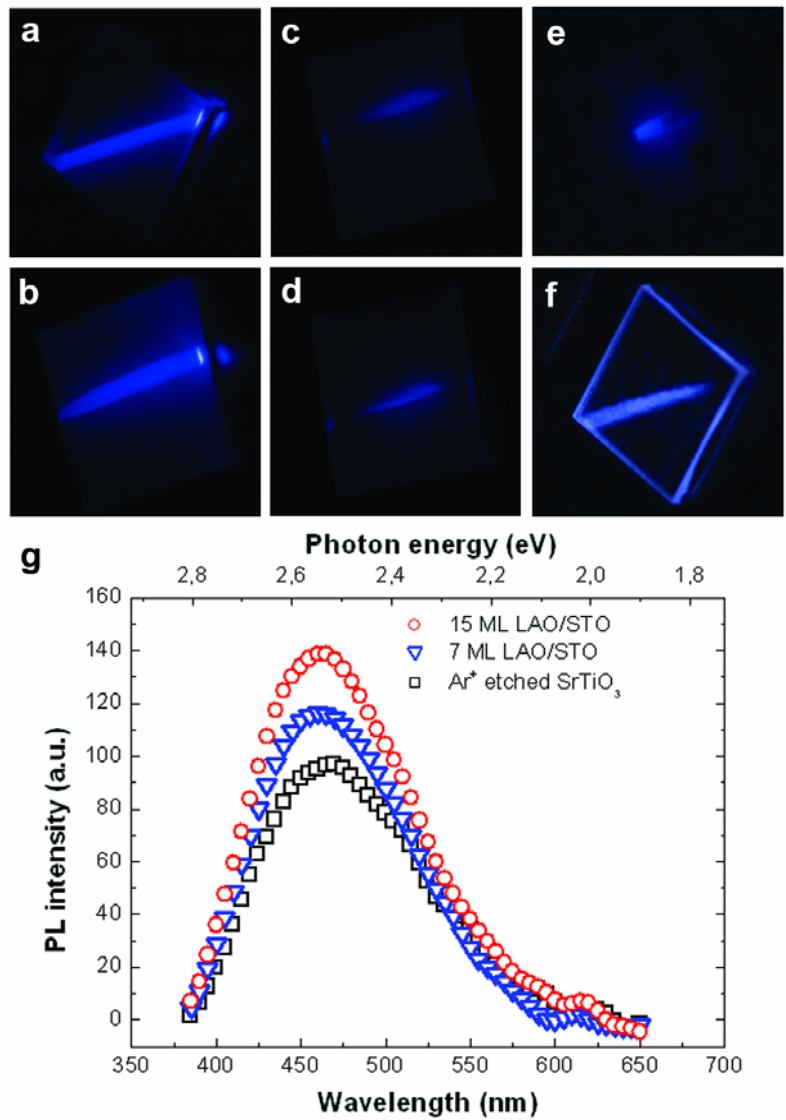

Figure 2. (a) – (f) Cathode luminescence from various SrTiO3 substrates and $LaAlO_3/SrTiO_3$-systems. (a) Substrate kept at $10^{-6}$ mbar without film deposition, and (b) after $LaAlO_3$ deposition. (c) Substrate kept at $10^{-4}$ mbar, and (d) after $LaAlO_3$ deposition. (e) Ar-ion irradiated $SrTiO_3$ single crystal substrate. (f) Cathode luminescence from $LaAlO_3$ substrate.

(g) Photo-luminescence measurements of $LaAlO_3/SrTiO_3$ films grown under low pressure conditions and of the Ar-ion irradiated $SrTiO_3$ substrate. Within the precision of the experiment the emitted light has the same peak wavelength (460 nm) for the three cases.

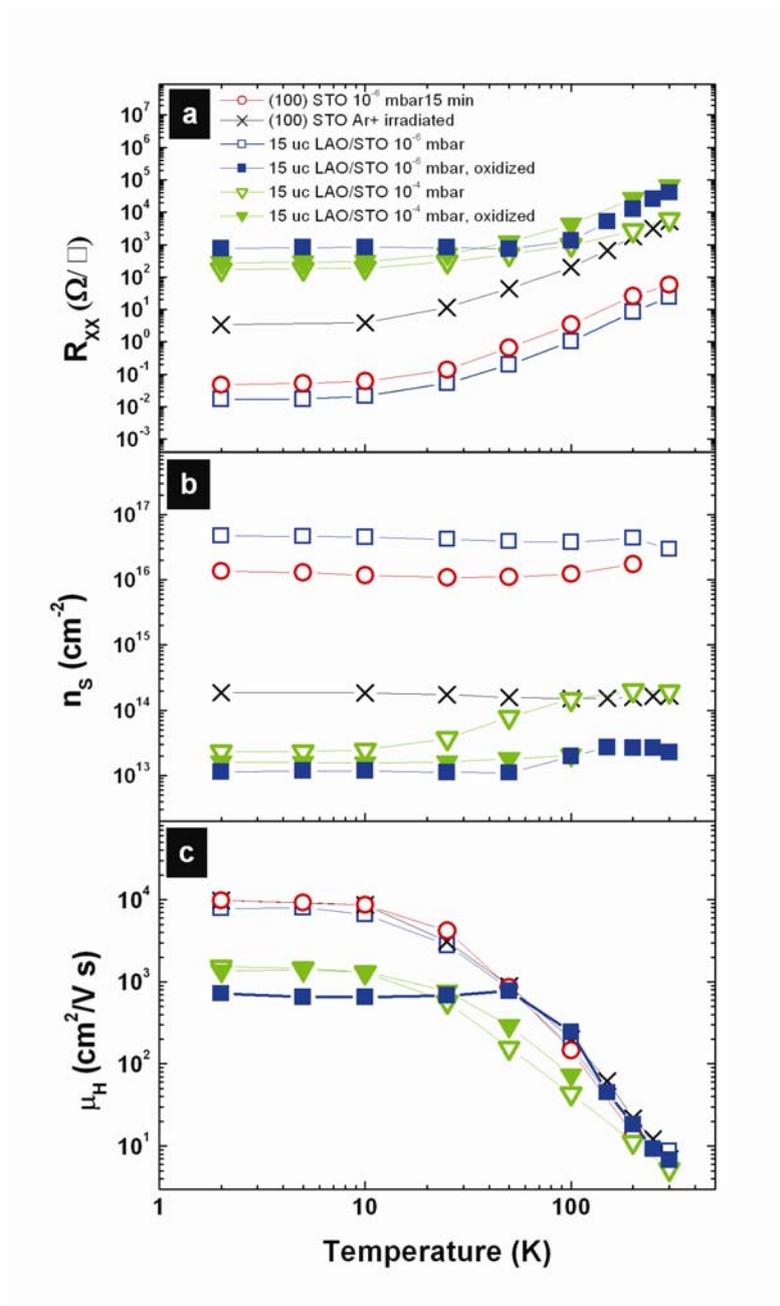

Figure 3. Sheet resistivity $R_{XX}$ (a), charge carriers density $n_S$ (b) and Hall mobility $\mu_H$ (c) for 15 u.c. LaAlO$_3$ film on SrTiO$_3$ substrate made at 10$^{-6}$ mbar oxygen pressure before and after oxidizing (open blue and solid rectangles), 15 u.c. LaAlO$_3$ film on SrTiO$_3$ substrate made at 10$^{-6}$ mbar oxygen pressure before and after oxidizing (open green and solid triangles), (100) SrTiO$_3$ substrate annealed at 10$^{-6}$ mbar for 15 min (open red circles) and Ar$^+$ -bombarded (100) SrTiO$_3$ substrate (black crosses).

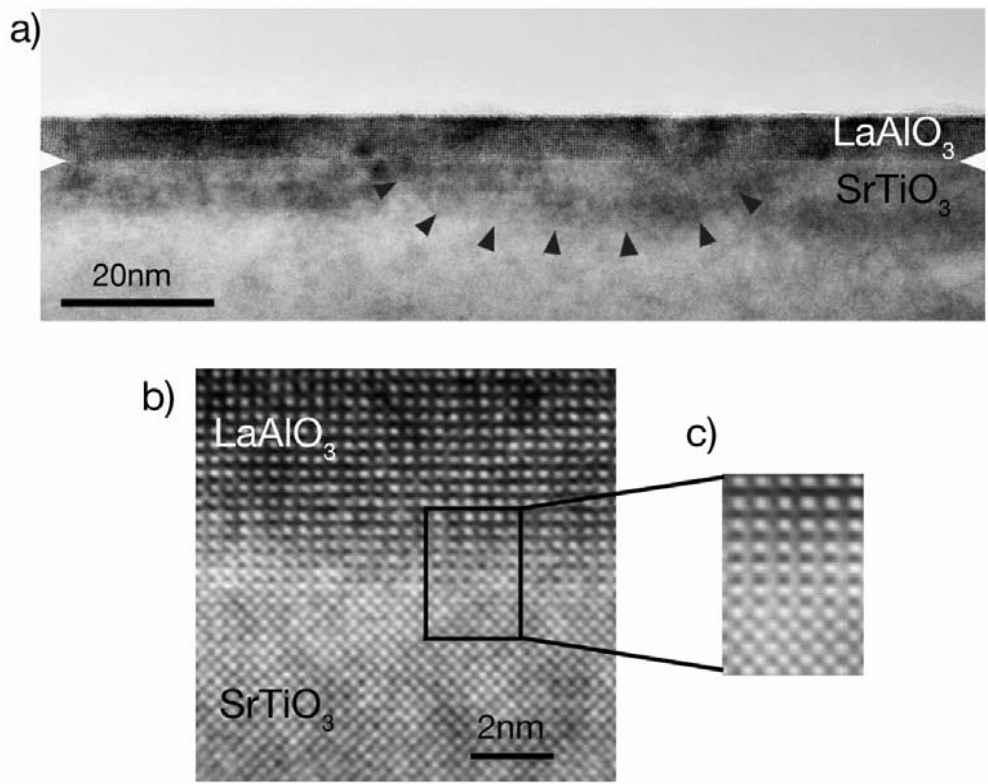

Figure 4. TEM cross section micrographs of the $LaAlO_3/SrTiO_3$ film deposited at oxygen pressure $10^{-6}$ mbar (a). U-shaped dark contrasts are marked with arrows. (b) The $LaAlO_3$ /$SrTiO_3$ interface imaged at higher magnification using high resolution microscopy showing details of the interface between the misfit dislocations. (c) A fast Fourier transform filtered image of the coherent interface between the dislocations.

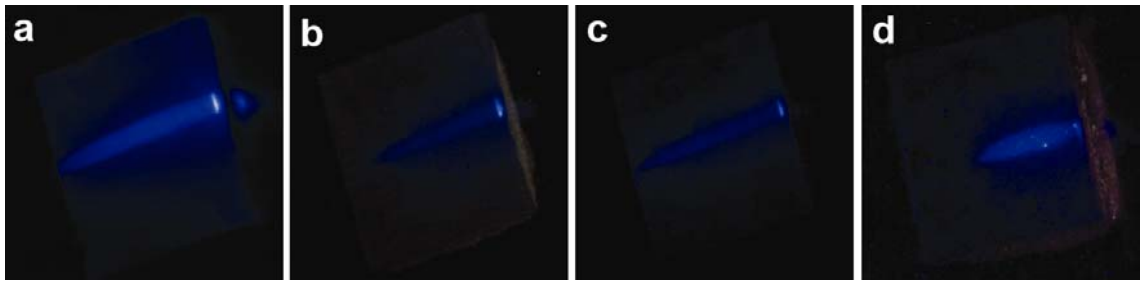

Figure 5. Cathode luminescence from a LaAlO$_3$/SrTiO$_3$ film (a) as deposited, (b) after re-oxygenation, (c) after vacuum annealing, and (d) after H$_2$/Ar annealing.